\documentclass[12pt,letterpaper]{article}
\usepackage[left=1in,top=1in,right=1in,nohead]{geometry}
\usepackage{enumerate}
\usepackage{graphicx}
\usepackage{amsmath}
\usepackage{amsthm}
\usepackage{mathrsfs}
\usepackage{amssymb}
\usepackage{multirow}
\usepackage{color}
\usepackage{bm}
\usepackage{comment}
\usepackage{subfigure}
\usepackage{paralist}
\usepackage{jheppub}

\theoremstyle{remark}
\newtheorem{thm}{Theorem}

\newtheorem{prop}{Proposition}
\newcommand{\grad}{\nabla}

\newcommand{\Ocal}{\mathcal{O}}
\newcommand{\eps}{\epsilon}
\newcommand{\Hcal}{\mathcal{H}}

\newcommand{\lambdat}{{\tilde{\lambda}}}

\newcommand{\kt}{\tilde{k}}
\newcommand{\lt}{\tilde{l}}
\newcommand{\xit}{\tilde{\xi}}
\newcommand{\gt}{\tilde{g}}
\newcommand{\gb}{\bar{g}}

\newcommand{\nt}{\tilde{n}}
\newcommand{\ttt}{\tilde{t}}
\newcommand{\intg}{\int_\gamma}

\DeclareMathOperator{\sech}{sech}

\def\bea#1\eea{\begin{align}#1\end{align}}
\def\be#1\ee{\begin{equation}#1\end{equation}}

\title{The Gravity Dual of Boundary Causality}

\author[a]{Netta Engelhardt}
\author[b]{and Sebastian Fischetti}

\affiliation[a]{Department of Physics, University of California, Santa Barbara, CA 93106, USA}
\affiliation[b]{Theoretical Physics Group, Blackett Laboratory, Imperial College, London SW7 2AZ, UK}

\emailAdd{engeln@physics.ucsb.edu}
\emailAdd{s.fischetti@imperial.ac.uk}

\abstract{In gauge/gravity duality, points which are not causally related on the boundary cannot be causally related through the bulk; this is the statement of boundary causality.  By the Gao-Wald theorem, the averaged null energy condition in the bulk is sufficient to ensure this property. Here we proceed in the converse direction: we derive a necessary as well as sufficient condition for the preservation of boundary causality under perturbative (quantum or stringy) corrections to the bulk.  The condition that we find is a (background-dependent) constraint on the amount by which light cones can ``open'' over all null bulk geodesics.  We show that this constraint is weaker than the averaged null energy condition.\\

}

\begin{document}

\maketitle
\flushbottom

\section{Introduction}
\label{sec:intro}

While fundamental properties of both quantum field theory and classical gravity are fairly well-understood, the basics of quantum gravity remain mysterious. A promising approach for shedding light on the latter is the AdS/CFT correspondence~\cite{Mal97, Wit98a, GubKle98}. This duality is expected to map fundamental principles of quantum field theory to equally essential properties of the gravitational dual. For instance, unitarity of the field theory implies that quantum string theory evolves unitarily\footnote{Similar in spirit but perhaps less intuitive are bulk consequences of laws of CFT entanglement, e.g.~\cite{LasMcD13, FauGui13, LasRab14, SwiVan14}.}. Causality is another pillar of quantum field theory; the principle that information cannot travel faster than light is as basic as unitarity.  The gauge/gravity dictionary must therefore translate causality into some equally important property of any regime of quantum gravity. What, then, is the gravitational bulk dual of boundary causality?

\begin{figure}[t]
\centering
\subfigure[]{
\includegraphics[page=1, width=0.25\textwidth]{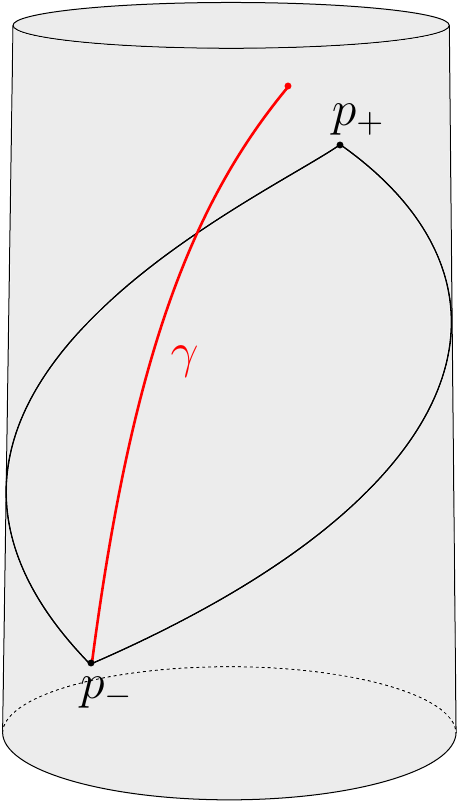}
\label{subfig:GaoWald}
}
\hspace{2cm}
\subfigure[]{
\includegraphics[page=2,width=0.25\textwidth]{Figures-pics}
\label{subfig:AdS}
}
\caption{\subref{subfig:GaoWald}: an illustration of a consequence of the Gao-Wald theorem~\cite{GaoWal00}. In any asymptotically AdS spacetime, lightlike signals fired along the boundary from the point $p_{-}$ reconverge at the point $p_{+}$. When the spacetime obeys the ANCC (and some other technical assumptions), every bulk null geodesic $\gamma$ (red) starting at $p_{-}$ arrives nowhere in the (causal) past of $p_{+}$.  \subref{subfig:AdS}: in an asymptotically AdS spacetime saturating the BCC (so that the spacetime saturates the ANCC and violates the null generic condition), there exists at least one null bulk geodesic $\gamma$ (red) connecting a pair of boundary points $p_{-}$ and $p_{+}$. Pure AdS saturates the BCC ``maximally'' in the sense that all null bulk geodesics arrive at the same time as their boundary-contained counterparts.  }
\label{fig:sketch}
\end{figure}

Significant progress towards answering this question was made by Gao and Wald in~\cite{GaoWal00}, who extended the earlier results of~\cite{Woo94}\footnote{See also~\cite{KleMcG01,PagSur02} for related work that derives necessary but not sufficient conditions on the bulk from boundary causality.}.  Besides imposing some technical assumptions, they required that the bulk obey the averaged null curvature condition (ANCC)\footnote{It is more common in the literature to invoke the Einstein field equation to replace~$R_{ab}k^a k^b$ in~\eqref{eq:ANCC} with~$T_{ab}k^a k^b$ and call the result the averaged null energy condition (ANEC)~\cite{Bor87}.  Here we wish to avoid imposing any dynamical equations to relate curvature to stress-energy, so we work with~\eqref{eq:ANCC} directly.}:
\be
\label{eq:ANCC}
\int_\gamma R_{ab} k^a k^b \, d\lambda \geq 0 \mbox{ for all } \gamma,
\ee
where~$R_{ab}$ is the Ricci tensor and the integral is taken over a complete null geodesic~$\gamma$ with affine parameter~$\lambda$ and tangent vector field~$k^a \equiv (\partial_\lambda)^a$.  Under such assumptions, Theorem~2 of~\cite{GaoWal00} showed that two boundary points are causally related through the bulk only if they are causally related on the boundary: there is no superluminal bulk signaling between boundary points (see Figure~\ref{subfig:GaoWald}).  We will refer to the requirement that this always be the case as the \textit{boundary causality condition} (BCC).  Note that as stated here, the BCC is a purely geometric condition, which is implied by the field-theoretic requirement of micro-causality (i.e.~that spacelike-separated operators commute); these geometric and field-theoretic notions of causality are \textit{a priori} distinct.

There is a glaring asymmetry here.  Causality is fundamental to QFT; any theory violating it is considered
unphysical and must be discarded.  The ANCC, however, is a reasonable assumption at the level of classical supergravity low energy limits of string theory, but it is not fundamental. Perfectly reasonable states of quantum fields can violate the ANCC already at the level of quantum field theory on a fixed classical background~\cite{Vis96,UrbOlu10}; violations have also been found when perturbative quantum corrections to the geometry are included~\cite{FlaWal96}. 

The reader may at this point protest that perturbative (quantum or stringy) corrections are unlikely to cause violations of the BCC.  Perturbative effects will only come at subleading order to classical supergravity contributions, and supergravity theories which are low-energy limits of string theory obey the ANCC.  Such bulk spacetimes will generically satisfy the BCC with a healthy margin: boundary points that are null separated through the bulk are timelike separated on the boundary, as shown in Figure~\ref{subfig:GaoWald}.  Thus at the perturbative level, corrections can only be significant when the spacetime in the classical gravity limit satisfies the BCC with no margin at all: there exist boundary points that are null separated through the bulk and also on the boundary, as shown in Figure~\ref{subfig:AdS}.  Therefore perturbative corrections only run into the danger of violating the BCC if the background spacetime simultaneously violates the null generic condition and saturates the ANCC. We know of only one spacetime with these properties: pure AdS itself, which has the feature that boundary points are null-related on the boundary if and only if they are null-related through the bulk. Conceivably, however, there could exist spacetimes in which only certain pairs of boundary points are null-related through both the bulk and boundary. These spacetimes would also saturate the BCC, though not as ``maximally'' as pure AdS does. In this paper, we will focus on the case of perturbations to pure AdS, though our calculations require only minor modifications for a generalization to other spacetimes which saturate the BCC.

More precisely, we take a converse approach to Gao-Wald: that is, we demand that a holographic spacetime (in the sense of AdS/CFT) preserve the causality of its QFT dual and investigate what restrictions this imposes on the gravitational side.  To this end, our assumptions on the bulk are minimal, and consist exclusively of the existence of a sharp and well-defined geometric causal structure generated by null geodesics\footnote{Note that this latter requirement may exclude, e.g.~Lovelock theories of gravity in the bulk, as in such theories the characteristic hypersurfaces need not be null~\cite{Ara87,Cho88,ReaTan14}.  We thank B.~Way for bringing this caveat to our attention.}. We find a remarkably simple result: roughly, a perturbation to AdS must cause light cones to close when averaged along any complete null geodesic. Crucially, this condition is weaker than the ANCC; that is, we will find an example of a perturbation which violates the ANCC but obeys our condition (and therefore still preserves boundary causality). 

We now give an intuitive explanation of this result, making it precise in the statement of Theorem~\ref{thm:main} below. We leave a detailed proof to Section~\ref{sec:result}, which can be safely skipped without loss of physical content. In Section~\ref{sec:ANCC} we reproduce a linearized version of the Gao-Wald theorem by showing that the ANCC implies the BCC.  As noted above, we show that the converse is false: our condition is strictly weaker than the ANCC.  We conclude in Section~\ref{sec:disc}.  Any technical proofs and calculations that do not add to the main discussion have been relegated to Appendices~\ref{app:basis} and~\ref{app:ANCC}.

\subsection{Explanation of Result}

\noindent Here we develop some intuition for our result. Consider a spacetime~$(M, \gb_{ab})$ whose metric~$\gb_{ab} = g_{ab} + \delta g_{ab}$ is a small perturbation of the pure AdS metric~$g_{ab}$.  In a perturbative analysis,~$g_{ab}$ and~$\gb_{ab}$ are simply different Lorentzian structures on the same manifold~$M$.  This feature allows the identification of points in the unperturbed and perturbed spacetimes, permitting a pointwise comparison of the causal structures induced by~$g_{ab}$ and~$\gb_{ab}$.  For instance, if~$p$ is a point in $M$ and~$k^a$ is null with respect to the AdS metric~$g_{ab}$, then the object
\be
\label{eq:pointwiseopening}
\gb_{ab} k^a k^b|_p = \delta g_{ab} k^a k^b|_p
\ee
measures the extent to which the perturbation ``opens up'' or ``closes'' the light cone at~$p$ in the~$k^a$ direction.  Since pure AdS saturates the BCC, we might roughly imagine that boundary causality will be preserved if the perturbation $\delta g_{ab}$ causes no light cones to open; that is, if~\eqref{eq:pointwiseopening} is non-negative everywhere for all vectors~$k^a$ which are null with respect to the background.

Of course, such a condition is highly gauge-dependent; as discussed in~\cite{GaoWal00}, it is possible to arbitrarily open or close the light cone of any point by an appropriate choice of an infinitesimal diffeomorphism~$\delta g_{ab} = 2\grad_{(a} \zeta_{b)}$.  However, as long the boundary causal structure is required to remain unchanged, such diffeomorphisms must vanish at the boundary.  This suggests that we might obtain a gauge-invariant quantity by integrating~\eqref{eq:pointwiseopening} along a complete null geodesic~$\gamma$ of the background AdS:
\be
\label{eq:averagedclosing}
I(\gamma) \equiv \int_\gamma \delta g_{ab} k^a k^b \, d\lambda,
\ee
with~$k^a = (\partial_\lambda)^a$ the affinely-parametrized tangent vector field to~$\gamma$.  It is straightforward to show that~$I(\gamma)$ vanishes for infinitesimal gauge transformations\footnote{More precisely, we will require (equation~\eqref{eq:falloffs}) that asymptotically~$\delta g_{ab} k^a k^b = \Ocal(\lambda^{-2})$, which implies that for a gauge transformation we must have~$\zeta_a k^a = \Ocal(\lambda^{-1})$.  This falloff is sufficient to guarantee that~\eqref{eq:averagedclosing} vanish.} and thus defines a gauge-invariant notion of the ``averaged'' light cone opening/closing along a given geodesic~$\gamma$.  We will therefore call~$I(\gamma)$ the \textit{averaged light cone tilt along~$\gamma$}.

An intuitive guess for a sufficient condition to preserve boundary causality would then be that~$\delta g_{ab}$ cause all light cones to close on average.  Indeed, we find this to be the case, as stated in our main Theorem:

\begin{thm}
\label{thm:main}
Let~$\delta g_{ab}$ be a regular and~$C^2$ perturbation of pure AdS$_{d+1}$ (with~$d \geq 2$) which leaves the causal structure of the boundary unchanged.  Define the averaged light cone tilt~$I(\gamma)$ along any complete null geodesic~$\gamma$ of pure AdS as in~\eqref{eq:averagedclosing}.  Then
\be
\label{eq:BCC}
I(\gamma) \geq 0 \mbox{ for all } \gamma
\ee
is both necessary and sufficient to ensure that boundary causality is preserved to linear order in~$\delta g_{ab}$.
\end{thm}

We emphasize that while our intuitive guess implied that~\eqref{eq:BCC} should be sufficient to satisfy the BCC, it is in fact a \textit{necessary} condition as well.  That is,~\eqref{eq:BCC} is equivalent to the BCC. The proof of this theorem, to which we now turn, involves requiring that bulk curves which are null and geodesic with respect to~$\gb_{ab}$ always reach the boundary to the future of their boundary-contained counterparts.


\section{Proof of Theorem \ref{thm:main}}
\label{sec:result}

Our conventions are as follows.  $\Hcal$ always refers to a Poincar\'e horizon of AdS, and~$\Sigma$ to its spatial geometry (which is shown below to be slicing-independent).  Undecorated objects ($g_{ab}$,~$R_{ab}$, etc.)~refer to the background AdS spacetime; objects dressed with bars ($\gb_{ab}$, etc.)~refer to the perturbed spacetime; objects dressed with tildes ($\gt_{ab}$, etc.)~refer to a conformal compactification of the background AdS:~$\gt_{ab} = \Omega^2 g_{ab}$.  Lower-case letters from the beginning of the Latin alphabet ($a$,~$b$,~$\ldots$) will be used as abstract indices; lower-case letters from the middle of the Latin alphabet ($i$,~$j$,~$\ldots$) will label elements of a spatial basis on~$\Sigma$; lower-case letters from the middle of the Greek alphabet ($\mu$,~$\nu$,~$\ldots$) will be used as coordinate indices for the global coordinates of AdS.  To emphasize the distinction between basis and abstract indices, we will explicitly write out any summations over basis indices.  Unspecified conventions are as in~\cite{Wald}.

\subsection*{Sketch of Proof} 

\begin{figure}[t]
\centering
\subfigure[]{
\includegraphics[page=3]{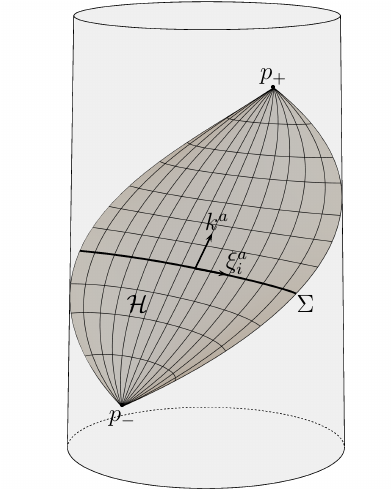}
\label{subfig:Poincare}
}
\subfigure[]{
\includegraphics[page=4]{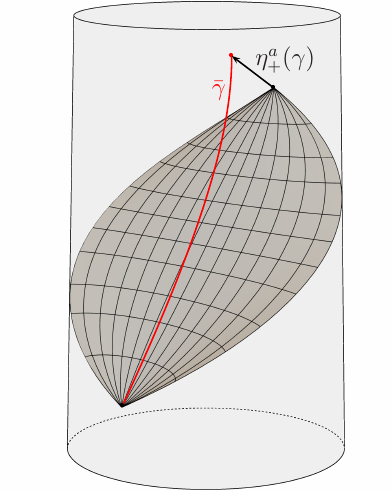}
\label{subfig:pert}
}
\caption{\subref{subfig:Poincare}: the boundary~$\Hcal$ of the future of a point~$p_-$ on the boundary of pure AdS; this surface is a Poincar\'e horizon of AdS.  Generators of~$\Hcal$ are shown as lines that emanate from~$p_-$ and reconverge at the antipodal boundary point~$p_+$.  Here we also highlight a spatial slice~$\Sigma$ (whose geometry is slice-independent), and we illustrate a basis adapted to~$\Hcal$.  This basis consists of the vector field~$k^a$ tangent to the generators of~$\Hcal$ as well as coordinate basis vectors~$\xi_i^a$. \subref{subfig:pert}: boundary causality requires that every null geodesic~$\bar{\gamma}$ (red) of the perturbed spacetime which starts at~$p_-$ must reach the boundary nowhere to the past of~$p_+$.  This requires that for all generators~$\gamma$ of~$\Hcal$, the deviation vector~$\eta^a_+(\gamma)$ cannot be past-directed.}
\label{fig:sketch}
\end{figure}

\noindent Consider the future~$J^+(p_-)$ of a point~$p_-$ on the AdS boundary; for pure AdS the boundary~$\dot{J}^+(p_-)$ of this future is just a Poincar\'e horizon~$\Hcal$ of AdS, as shown in Figure~\ref{subfig:Poincare}.  The generators of~$\Hcal$ are null and geodesic with respect to the AdS metric~$g_{ab}$, and they reconverge at the boundary point~$p_+$.  In the perturbed spacetime~$\gb_{ab}$, however, the boundary~$\dot{J}^+_\mathrm{pert}(p_-)$ will be generated by curves $\bar{\gamma}$ that are null and geodesic with respect to~$\gb_{ab}$.  Boundary causality is then the requirement that none of the generators of~$\dot{J}^+_\mathrm{pert}(p_-)$ intersect the boundary to the past of~$p_{+}$.  Since~$\delta g_{ab} \equiv \gb_{ab} - g_{ab}$ is infinitesimal, we may understand this requirement better by considering the deviation vector field~$\eta^a$ between generators of~$\Hcal$ and generators of~$\dot{J}^+_\mathrm{pert}(p_-)$; we can construct~$\eta^a$ over all of~$\Hcal$ generator-by-generator.  Because the generators of~$\Hcal$ converge at the boundary points~$p_-$ and~$p_+$,~$\eta^a$ is ill-defined there.  However,~$\eta^a$ admits well-defined limits as~$p_\pm$ are approached along individual generators~$\gamma$.  We will use~$\eta^a_\pm(\gamma)$ to denote these ($\gamma$-dependent) limiting values.  By construction, we will need~$\eta^a_-(\gamma)$ to vanish, while boundary causality will be preserved if and only if~$\eta^a_+(\gamma)$ is never past-directed for any~$\gamma$, as shown in Figure~\ref{subfig:pert}.  Requiring this to hold for all boundary points~$p_-$ (and therefore all Poincar\'e horizons~$\Hcal$), we obtain precisely condition~\eqref{eq:BCC}, completing the proof.

\subsection*{Detailed Proof}

In order to solve for the deviation vector~$\eta^a$, we will begin with a useful observation:

\begin{prop}
\label{prop:Poincare}
Let~$\Hcal$ be a Poincar\'e horizon of pure AdS$_{d+1}$ (with~$d \geq 2$).  Let~$\lambda$ be an affine parameter along the generators of~$\Hcal$, and label these generators by~$(d-1)$ parameters~$s^i$.  Then (i) the intrinsic spatial geometry~$\Sigma$ of~$\Hcal$ is slicing-independent (i.e.~independent of~$\lambda$), and (ii) the coordinate basis vector fields~$\xi_i^a \equiv (\partial_{s^i})^a$ can be taken to be parallel transported along the tangent vector field~$k^a \equiv (\partial_\lambda)^a$, and therefore along the generators of~$\Hcal$.
\end{prop}

The proof of this Proposition, as well as an explicit construction of such a coordinate system, are left to Appendix~\ref{app:basis}.  Here we note that property~(i) is a consequence only of the fact that Poincar\'e horizons are Killing horizons, while property~(ii) also requires use of the fact that AdS is maximally symmetric (specifically, that its Riemann tensor can be written as~\eqref{eq:riemann} below).

Now consider some Poincar\'e horizon~$\Hcal$ with spatial geometry~$\Sigma$; by Proposition~\ref{prop:Poincare}, we may introduce a basis everywhere on~$\Hcal$ which is parallel-transported along $k^{a}$ and consists of~$k^a$,~$\xi_i^a$, and an auxiliary null vector field~$l^a$. We take $l^{a}$ to be orthogonal to the~$\xi_i^a$ and normalized so that~$l \cdot k = 1$ (also note that unlike~$k^a$ and~$\xi_i^a$, we do not take~$l^a$ to be a coordinate basis vector).  The components of the induced metric on~$\Sigma$ are given by
\be
q_{ij} = \xi_i^a \xi_j^b g_{ab} = \xi_i \cdot \xi_j,
\ee
and the full spacetime metric can then be written as
\be
g_{ab} = 2k_{(a} l_{b)} + \sum_{i,j} q^{ij} (\xi_i)_a (\xi_j)_b,
\ee
where~$q^{ij}$ is the matrix inverse of~$q_{ij}$.

We are now prepared to solve for the deviation vector field~$\eta^a$ everywhere on~$\Hcal$.  This deviation vector field will obey the inhomogeneous equation of geodesic deviation (also called the generalized Jacobi equation)~\cite{PynBir93} along every generator:
\be
\label{eq:geodeviation}
k^a \grad_a \left(k^b \grad_b \eta^c\right) + R_{abd}^{\phantom{abd}c} k^a k^d \eta^b = -\delta \Gamma^c_{ab} k^a k^b,
\ee
where
\be
\delta \Gamma^a_{bc} = \frac{1}{2} \, g^{ad} \left(\grad_b \delta g_{cd} + \grad_c \delta g_{bd} - \grad_d \delta g_{bc}\right),
\ee
and~$R_{abcd}$ is the Riemann tensor of pure AdS, which can be written as
\be
\label{eq:riemann}
R_{abcd} = \frac{2}{\ell^2}\, g_{a[d} g_{c]b},
\ee
with~$\ell$ the AdS scale.  The vector field tangent to the generators of~$\dot{J}_\mathrm{pert}^+(p)$ will be
\be
\bar{k}^a \equiv k^a + \delta k^a = k^a + k^b \grad_b \eta^a.
\ee
Requiring that~$\bar{k}^a$ be null with respect to~$\gb_{ab}$, we obtain an equation for the scalar component~$k \cdot \eta$:
\be
\label{eq:constraint}
\gb_{ab} \bar{k}^a \bar{k}^b = 0 \Rightarrow k^a \grad_a (k \cdot \eta) = -\frac{1}{2} \delta g_{ab} k^a k^b.
\ee
This constraint is preserved by the deviation equation, as can be checked by contracting~\eqref{eq:geodeviation} with~$k^a$.

The contraction of~\eqref{eq:geodeviation} with the remaining basis vectors~$l^a$ and~$\xi_i^a$ implies via~\eqref{eq:riemann}:
\begin{subequations}
\label{eqs:etacomps}
\bea
k^a \grad_a \left(k^b \grad_b (l \cdot \eta)\right) &= -\frac{1}{\ell^2} \left(k \cdot \eta\right) - \delta \Gamma^c_{ab} k^a k^b l_c, \\
k^a \grad_a \left(k^b \grad_b (\xi_i \cdot \eta)\right) &= -\delta \Gamma^c_{ab} k^a k^b (\xi_i)_c.
\eea
\end{subequations}
Equations~\eqref{eq:constraint} and~\eqref{eqs:etacomps} are linear ordinary differential equations for the components of~$\eta^a$ in the adapted basis~$(k^a, l^a, \xi_i^a)$.  These can be solved by direct integration; after integrations by parts, they yield
\begin{subequations}
\label{eqs:etasolution}
\be
k\cdot \eta = a - \frac{1}{2} \int_{\lambda_0}^\lambda \delta g_{ab} k^a k^b \, d\lambda',
\ee
\begin{multline}
l \cdot \eta = b + c \lambda - \frac{a}{2\ell^2} \, \lambda^2 + \frac{1}{4\ell^2} \int_{\lambda_0}^\lambda (\lambda - \lambda')^2 \delta g_{ab} k^a k^b \, d\lambda' \\ - \int_{\lambda_0}^\lambda \delta g_{ab} k^a l^b \, d\lambda' + \frac{1}{2} \int_{\lambda_0}^\lambda (\lambda-\lambda')  k^a k^b l^c \grad_c \delta g_{ab} \, d\lambda',
\end{multline}
\be
\xi_i \cdot \eta = d_i + e_i \lambda - \int_{\lambda_0}^\lambda \delta g_{ab} k^a \xi_i^b \, d\lambda' + \frac{1}{2} \int_{\lambda_0}^\lambda (\lambda-\lambda') \xi_i^c \grad_c (\delta g_{ab} k^a k^b) \, d\lambda',
\ee
\end{subequations}
where~$a$,~$b$,~$c$,~$d_i$, and~$e_i$ are constants of integration and~$\lambda_0$ is an arbitrary reference affine parameter (all of which may depend on the spatial coordinates~$s^i$).

To fix some of the integration constants we must impose appropriate boundary conditions: namely, that the generators of~$\Hcal$ and~$\dot{J}^+_\mathrm{pert}(p_-)$ originate at the same boundary point~$p_-$, and thus that
\be
\label{eq:minusbndry}
\eta^a_-(\gamma) \equiv \underset{\mathrm{on} \, \gamma}{\lim_{\lambda \to -\infty}} \eta^a=0
\ee
for every generator~$\gamma$ (corresponding to fixed~$s^i$).  Because the metric~$g_{ab}$ and the basis~$(k^a, l^a, \xi_i^a)$ are singular at the boundary, it is much simpler to work in a geometry that is regular there.  To this end, consider the conformally compactified geometry~$\gt_{ab} \equiv \Omega^2 g_{ab}$,~$\widetilde{\delta g}_{ab} \equiv \Omega^2 \delta g_{ab}$ where~$\Omega$ is some conformal factor which renders $\gt_{ab}$ regular at the boundary.  The rescaled basis
\be
\label{eq:tildebasis}
\left(\kt^a, \lt^a, \xit^a_i\right) = \left(\frac{1}{\Omega^2} \, k^a, l^a, \frac{1}{\Omega} \, \xi^a_i\right)
\ee
is regular at the boundary with respect to~$\gt_{ab}$ (since~$\kt^a$ is geodesic and affinely-parametrized with respect to~$\gt_{ab}$; see, e.g.~Appendix~D of~\cite{Wald}).  The requirement that~$\delta g_{ab}$ maintain the causal structure of the boundary implies that the components of~$\widetilde{\delta g}_{ab}$ in the basis~\eqref{eq:tildebasis} must be finite\footnote{In fact, if they are nonzero, they must induce a conformal transformation of the boundary metric.  We will not use this information here.}.  Noting that along any geodesic~$\gamma$ we must have asymptotically~$\Omega \sim 1/\lambda + \cdots$\footnote{This can be seen easily for pure AdS from, e.g.~the explicit parametrizations~\eqref{eq:Poincaregeodesic}, though it holds generally in any asymptotically locally AdS spacetime.}, from this requirement we deduce the asymptotic falloffs
\be
\label{eq:falloffs}
\delta g_{ab} k^a k^b = \Ocal(\lambda^{-2}), \qquad \delta g_{ab} k^a l^b = \Ocal(\lambda^0), \qquad \delta g_{ab} k^a \xi_i^b = \Ocal(\lambda^{-1}).
\ee

We can now impose the boundary condition~\eqref{eq:minusbndry}.  In the compactified basis~\eqref{eq:tildebasis}, the components of the deviation vector~$\eta^a$ are
\be
\label{eq:etatilde}
\widetilde{k \cdot \eta} \equiv \gt_{ab} \kt^a \eta^b = k \cdot \eta, \quad \widetilde{l \cdot \eta} \equiv \gt_{ab} \lt^a \eta^b = \Omega^2 \, l \cdot \eta, \quad \widetilde{\xi_i \cdot \eta} \equiv \gt_{ab} \xit_i^a \eta^b = \Omega \, \xi_i \cdot \eta.
\ee
These must vanish at~$p_-$; comparing with~\eqref{eqs:etasolution} and using the falloffs~\eqref{eq:falloffs}, we find
\be
\label{eq:bndryconds}
a = -\frac{1}{2} \int_{-\infty}^{\lambda_0} \delta g_{ab} k^a k^b \, d\lambda', \quad e_i = \frac{1}{2} \int_{-\infty}^{\lambda_0} \xi_i^c \grad_c (\delta g_{ab} k^a k^b) \, d\lambda'.
\ee
The remaining unspecified constants~$b$,~$c$,~$d_i$ (which correspond to solutions to the homogeneous equation of geodesic deviation) represent the freedom to shift and rescale the affine parameter~$\lambda$ or to change the initial momentum of the geodesic, respectively.

Plugging~\eqref{eq:bndryconds} back into~\eqref{eqs:etasolution} and again using~\eqref{eq:etatilde}, we can compute the components of the deviation vector $\eta^{a}_{+}(\gamma)$ at $p_{+}$:
\begin{subequations}
\label{eqs:nulltotcomponents}
\bea
\widetilde{k \cdot \eta}_+(\gamma) \equiv \underset{\mathrm{on} \, \gamma}{\lim_{\lambda \to \infty}} \widetilde{k \cdot \eta} &= - \frac{1}{2} \int_\gamma \delta g_{ab} k^a k^b \, d\lambda = -\frac{1}{2} I, \label{eq:etak} \\
\widetilde{l \cdot \eta}_+(\gamma) \equiv \underset{\mathrm{on} \, \gamma}{\lim_{\lambda \to \infty}} \widetilde{l \cdot \eta} &= \frac{\omega^2}{4\ell^2} \int_\gamma \delta g_{ab} k^a k^b \, d\lambda = \frac{\omega^2}{4\ell^2} I, \label{eq:etaell} \\
\widetilde{\xi_i \cdot \eta}_+(\gamma) \equiv \underset{\mathrm{on} \, \gamma}{\lim_{\lambda \to \infty}} \widetilde{\xi_i \cdot \eta} &= \frac{\omega}{2} \int_\gamma  \xi_i^c \grad_c (\delta g_{ab} k^a k^b) \, d\lambda = \frac{\omega}{2} D_i I,  \label{eq:etai}
\eea
\end{subequations}
where~$\omega \equiv \lim_{\lambda \to \infty \, \mathrm{on} \, \gamma} \Omega \lambda$ and~$D_i$ is the covariant derivative on~$\Sigma$ (so~$D_i I = \partial I/\partial s^i$); recall that $I$ was defined  in~\eqref{eq:averagedclosing}.  We have left all~$\gamma$-dependence implicit in quantities on the right-hand side.  We note that since~$I$ is defined for every generator~$\gamma$ of~$\Hcal$, and since every such generator corresponds to a unique point in~$\Sigma$, we are now treating~$I$ as a scalar on~$\Sigma$.

Expanding~$\eta^a_+(\gamma)$ in the tilded basis and using~\eqref{eqs:nulltotcomponents} we find that
\be
\label{eq:etabndy}
\eta^a_+(\gamma) = \frac{\omega}{2\ell}\left[\left(\frac{\omega}{2\ell} \kt^a - \frac{\ell}{\omega} \lt^a\right) I + \ell \sum_{i,j} q^{ij} \xit_i^a D_j I \right],
\ee
where here the tilded basis vectors are understood to be evaluated as~$\lambda \to \infty$ along the generator~$\gamma$.  Note that the vector~$\ttt^a \equiv (\omega/2\ell) \kt^a - (\ell/\omega) \lt^a$ is a timelike and future-directed unit vector (with respect to~$\gt_{ab}$)\footnote{In fact, it is possible to show that~$\ttt^a$ is the projection of~$\kt^a$ onto the boundary:~$\ttt^a = (\omega/\ell) (\delta^a_{\phantom{a}b} - \nt^a \nt_b) \kt^b$, where~$\nt^a$ is the unit normal (with respect to~$\gt_{ab}$) to the boundary.}.  Since it is also orthogonal to the~$\xit_i^a$, equation~\eqref{eq:etabndy} is a decomposition of~$\eta^a_+(\gamma)$ in a Lorentz frame specified by~$\ttt^a$ and the subspace spanned by the~$\xit_i^a$.  Requiring that~$\eta^a_+(\gamma)$ not be past-directed yields
\be
\label{eq:condition}
I \geq -\ell \sqrt{\sum_{i,j} q^{ij} (D_i I)(D_j I)},
\ee
with equality only if both sides vanish. This is precisely the condition that boundary causality be preserved on~$\Hcal$ generator-by-generator.  Requiring $I\geq 0$ along all null geodesics is therefore a sufficient condition for~\eqref{eq:condition} to be obeyed; this proves sufficiency in Theorem~\ref{thm:main}.

To prove necessity, we proceed by contradiction.  Assume that there exists a metric perturbation that satisfies~\eqref{eq:condition} on all null geodesics, but such that there exists a geodesic~$\gamma$ on which~$I < 0$.  Let~$\Hcal$ be the Poincar\'e horizon to which~$\gamma$ belongs, and as usual let~$\Sigma$ be its spatial geometry.

Recall that~$I$ is a scalar on~$\Sigma$.  The boundary~$\partial \Sigma$ of~$\Sigma$ corresponds to those null geodesics lying entirely on the AdS boundary.  Since the boundary causal structure is unchanged, these geodesics are unperturbed, and so~$I(\partial \Sigma) = 0$.  By assumption, we also have that~$I$ becomes negative somewhere in~$\Sigma$.  Thus since~$I$ must vanish on~$\partial \Sigma$, and since~$I$ is~$C^2$ (because $\delta g_{ab}$ is $C^{2}$), there must be some (global) minimum at which~$I < 0$ and~$D_i I = 0$.  But then~\eqref{eq:condition} is violated at this minimum, in contradiction with the assumption that it be obeyed everywhere.  This proves necessity, and thus completes the proof.


\section{Connection to Gao-Wald}
\label{sec:ANCC}

As noted in Section~\ref{sec:intro}, the Gao-Wald theorem provides a solely sufficient condition for preserving boundary causality: the ANCC.  It would thus be very illuminating to understand the relationship between our condition and the ANCC.  Pure AdS is Einstein and therefore saturates the ANCC; to compute the averaged null curvature for linearized perturbations of AdS, we need only find the first-order variation of~\eqref{eq:ANCC} above.  The details are unilluminating, so we relegate them to Appendix~\ref{app:ANCC}.  The result is
\be
\label{eq:linANCC}
\delta \left(\intg R_{ab} k^a k^b \, d\lambda \right) = -\frac{1}{2} \left(D^2 I - \frac{d-1}{\ell^2} \, I\right),
\ee
where~$I$ is as defined in~\eqref{eq:averagedclosing} and~$D^2 = \sum_{ij} q^{ij} D_i D_j$ is the (covariant) Laplacian on~$\Sigma$, the spatial geometry of the Poincar\'e horizon to which~$\gamma$ belongs.

This result allows us to prove the following Theorem:
\begin{thm}
Let~$\delta g_{ab}$ be a regular and~$C^2$ perturbation of pure AdS$_{d+1}$ (with~$d \geq 2$) which leaves the causal structure of the boundary unchanged.  Then the ANCC implies the BCC, but the BCC does not imply the ANCC.  The ANCC is thus sufficient but not necessary for preserving boundary causality. \textcolor{white}{FIRMLY.}
\end{thm}

That the ANCC is sufficient to ensure boundary causality is of course just the Gao-Wald theorem; the proof in the linear regime serves as a check of our results.  More interesting is the fact that the ANCC is not necessary: as expected, this implies that our boundary causality condition is a genuinely new requirement on the bulk geometry, and is weaker than the ANCC.  We now prove these statements.

\begin{proof}
Sufficiency is immediately implied by Gao-Wald.  Here we provide an additional proof in the linearized regime to make contact with our formalism.  Proceed by contradiction: assume that the ANCC is obeyed to linear order in~$\delta g_{ab}$ but that there exists a geodesic~$\gamma$ of pure AdS on which the BCC~$I(\gamma) \geq 0$ is violated.  Let~$\Hcal$ (with associated spatial geometry~$\Sigma$) be the Poincar\'e horizon to which~$\gamma$ belongs; then from~\eqref{eq:linANCC}, the ANCC implies
\be
\label{eq:ANCCproof}
D^2 I - \frac{d-1}{\ell^2} \, I \leq 0
\ee
everywhere on~$\Sigma$.  Next, recall that as discussed in the proof of Theorem~\ref{thm:main}, we have~$I(\partial \Sigma) = 0$.  By assumption, we also have that~$I < 0$ somewhere on~$\Sigma$.  Since~$I$ is~$C^2$ (since~$\delta g_{ab}$ is), these two requirements imply that there must exist some local minimum at which~$I < 0$ and~$D^2 I \geq 0$.  But the existence of such a minimum is a contradiction, as it would violate~\eqref{eq:ANCCproof}.  Thus~$I \geq 0$ everywhere, which proves sufficiency.

To prove that the BCC does not imply the ANCC, we provide an example of a perturbation that obeys the former but violates the latter.  To that end, consider the  perturbation which, when expressed in the global coordinates of AdS (given in~\eqref{eq:AdSglobal}), takes the form
\be
\sum_{\mu,\nu} \delta g_{\mu\nu} \, dx^\mu \, dx^\nu = \frac{\eps \, r^2/\ell^2}{(1+r^2/\ell^2)^d} \, d\ttt^2
\ee
with~$\eps > 0$.  This perturbation is static and spherically symmetric, and can be thought of as an attractive gravitational potential.  By converting to the coordinates~\eqref{eq:adaptedmetric} adapted to the Poincar\'e horizon~$\tau = 0$, it is straightforward to compute the integral for~$I$ via the residue theorem.  We find
\be
I = A \left((2d+1)\tanh^2\chi+1\right)\sech^{2d-1}\chi
\ee
for some positive constant~$A$\footnote{Explicitly,
\[
A = \eps \ell \frac{\sqrt{\pi}}{4} \, \frac{\Gamma(d+1/2)}{\Gamma(d+2)}.
\]}, where~$\chi \in [0,\infty)$ is a coordinate on~$\Sigma$.  Since~$I$ is positive for all~$\chi$, these perturbations obey the BCC.  However, we find that
\be
D^2 I - \frac{d-1}{\ell^2} I = \frac{2^{2d+1} d^2(d+1)}{\ell^2} \, A \, e^{-(2d-1)\chi}\left[1 + \Ocal(e^{-\chi})\right],
\ee
so this perturbation violates the linearized ANCC~\eqref{eq:ANCCproof} at large~$\chi$.  Thus the linearized ANCC is not a necessary condition to ensure boundary causality.
\end{proof}

\section{Discussion}
\label{sec:disc}

We have shown that the preservation of boundary causality is perturbatively equivalent to the averaged light cone tilt condition~\eqref{eq:BCC} in the bulk.  We have furthermore shown that this condition is strictly weaker than the ANCC assumed in the Gao-Wald theorem.  Moreover, it is worth noting that the ANCC is itself one of the weakest assumptions typically made of the bulk geometry.  We emphasize that our results show that even this assumption is stronger than necessary for a construction of the bulk guided by this physical requirement of the CFT.

One obvious drawback of our condition is its manifest background-dependence: it is explicitly formulated in the context of linear perturbations about pure AdS.   As emphasized in Section~\ref{sec:intro}, such perturbations
 of pure AdS are the primary ones of physical relevance, as AdS is the only spacetime of which we are aware that saturates the BCC\footnote{We repeat that our analysis applies with little modification to other spacetimes which saturate the BCC, should they exist.}.  Nevertheless, this background-dependence is undesirable for reasons both aesthetic and conceptual: while~\eqref{eq:BCC} must be a fundamental constraint on the bulk geometry, it should presumably arise as the linearization of some background-independent principle in perturbative quantum gravity. The explicit background-dependence of our result thus hinders an understanding of its broader consequences. It would therefore  be interesting to construct a condition which yields~\eqref{eq:BCC} when linearized around pure AdS. 
 
Our analysis was motivated by consideration of perturbative (quantum or stringy) corrections to the bulk geometry.  Now, at sufficiently high orders in quantum corrections, the geometry  undergoes fluctuations, resulting in a dynamically fluctuating causal structure.  This may prima facie suggest that our analysis is only applicable to stringy corrections due to our assumption of a sharp causal structure.  However, this objection is unfounded:  there certainly exist quantum corrections to pure AdS which at leading order will leave the causal structure sharp. Note that other quantum corrections will only induce perturbatively small ``fuzziness'' in the causal structure, possibly suggesting that there may be a ``fuzzy'' version of our result (perhaps similar in spirit to the smeared-out ANCC of~\cite{FlaWal96}).

\begin{figure}[t]
\centering
\includegraphics[page=6]{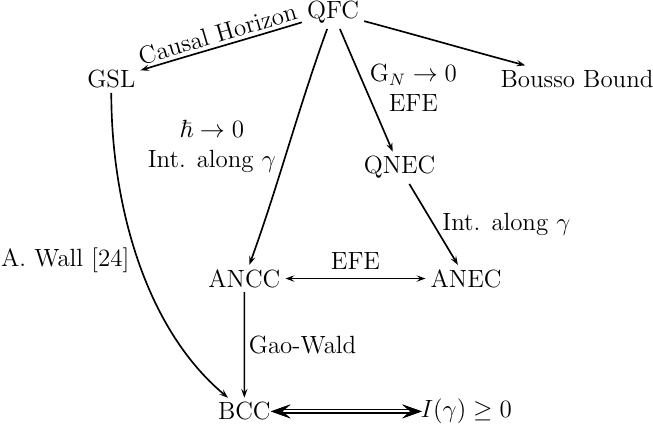}
\caption{A flowchart showing the logical relationships between assorted conditions often assumed in perturbative and semiclassical gravity.  Acronyms are defined in the main text, with the exception of the Einstein field equation (EFE).  Ignoring assorted assumptions, the implications are as follows: the QFC implies the GSL when applied to a causal horizon, which in turn was shown in~\cite{Wal10QST} to imply the BCC.  The QFC also implies the ANCC in the classical limit ($\hbar \to 0$) when integrated along a complete null geodesic, which in turn likewise implies the BCC by Gao-Wald.  In the probe limit ($G_{N}\rightarrow 0$) and under use of the EFE, the QFC implies the QNEC, which in turn implies the ANEC when integrated along complete null geodesics; the ANCC and the ANEC are equivalent when the EFE are invoked.  We emphasize that in this paper we are not restricting ourselves to either the~$\hbar \to 0$ or~$G_N \to 0$ limits.  Finally, we note that the QFC also implies the Bousso bound~\cite{Bou99b,Bou99c,Bou02}, but it is not clear if and how this bound is related to the BCC.}
\label{fig:conditions}
\end{figure}

Other conditions on perturbative quantum gravity are limited in number, so it is natural to ask whether our condition is related to those previously explored in the literature.  We will not attempt to answer this question in this paper.  However, in Figure~\ref{fig:conditions} we illustrate the connections between the BCC and reasonable assumptions about the bulk that have been shown to guarantee boundary causality.  The conditions in the figure all stem from the Quantum Focussing Conjecture (QFC)~\cite{BouFis15}, which is the strongest of the ones listed; these include the Generalized Second Law (GSL), the Quantum Null Energy Condition (QNEC)~\cite{BouFis15, QNEC,KoeLei15}, and the ANCC.  Most of these involve the generalized entropy, a quantity nonlocal in space and local in time.  This is in contrast with our condition, which is nonlocal in time and local in space, and which we have shown is weaker than the ANCC. This implies that our condition is weaker than any of those listed in the central branch of Figure~\ref{fig:conditions}; it would be interesting to determine the relationship between our condition and the GSL. \textcolor{white}{FIN.}

\section*{Acknowledgements}

It is a pleasure to thank Dalit Engelhardt, Gary Horowitz, Don Marolf, Mark Van Raamsdonk, and Benson Way for useful discussions; we thank John Hammond for inspiration.  SF also wishes to thank the University of California, Santa Barbara for hospitality at early and late stages of this work.  NE was supported in part by NSF grant PHY-1504541, the NSF Graduate Research Fellowship under grant DE-1144085, and by funds from the University of California. SF was supported by the ERC Advanced grant No.~290456.
\appendix

\section{Proof of Proposition~\ref{prop:Poincare}}
\label{app:basis}

In this Appendix, we prove Proposition~\ref{prop:Poincare}:
\setcounter{prop}{0}
\begin{prop}
Let~$\Hcal$ be a Poincar\'e horizon of pure AdS$_{d+1}$ (with~$d \geq 2$).  Let~$\lambda$ be an affine parameter along the generators of~$\Hcal$, and label these generators by~$(d-1)$ parameters~$s^i$.  Then (i) the intrinsic spatial geometry~$\Sigma$ of~$\Hcal$ is slicing-independent (i.e.~independent of~$\lambda$), and (ii) the coordinate basis vector fields~$\xi_i^a \equiv (\partial_{s^i})^a$ can be taken to be parallel transported along the tangent vector field~$k^a \equiv (\partial_\lambda)^a$, and therefore along the generators of~$\Hcal$.
\end{prop}
We will provide two different proofs.  In Section~\ref{subapp:general}, we use only general properties of Killing horizons and the fact that the bulk is pure AdS.  In Section~\ref{subapp:explicit}, we will provide a constructive proof by explicitly writing the metric of AdS in such an adapted coordinate system.

\subsection{General Proof}
\label{subapp:general}

Statement~(i) is quite intuitive, as~$\Hcal$ is a Killing horizon and thus evolution along it is an isometry of the spacetime.  To prove it, recall that in the coordinate basis specified by the~$s^i$, the components of the induced metric on~$\Sigma$ are
\be
q_{ij} = \xi_i^a \xi_j^b g_{ab} = \xi_i \cdot \xi_j.
\ee
To show that~$\Sigma$ is slicing-independent, we show that the components~$q_{ij}$ are independent of~$\lambda$.  To see this, recall that since~$\Hcal$ is a Killing horizon, its expansion, shear, and twist all vanish (see e.g.~Chapter~12.5 of~\cite{Wald}).  Thus the contraction of the tensor~$B_{ab} = \grad_a k_b$ with any vector contained in~$\Hcal$ must be proportional to~$k_a$ (this follows because any such contraction must be in the equivalence class of vectors that differ from~$k^a$ by zero; see Chapter~9 of~\cite{Wald}).  But since the coordinate vectors~$\xi_i^a$ are deviation vectors\footnote{Here we remind the reader that in this section these are deviation vectors between generators of~$\Hcal$, and not between generators of~$\Hcal$ and~$\dot{J}^+_\mathrm{pert}(p_-)$.} within~$\Hcal$, they are contained in~$\Hcal$, and so we must have
\be
\label{eq:xigradk}
\xi_i^a \grad_a k^b = \alpha_i k^b
\ee
for some functions~$\alpha_i$.  Furthermore, since~$k^a$ and~$\xi_i^a$ are coordinate basis vector fields, their commutator vanishes, and thus
\be
\label{eq:kxicomm}
\left[k,\xi_i\right]^a = 0 \Rightarrow k^b \grad_b \xi_i^a = \alpha_i k^a.
\ee
Then using the fact that the~$\xi_i^a$ are orthogonal to~$k^a$ (since~$k^a$ is orthogonal to any vector in~$\Hcal$),~\eqref{eq:kxicomm} implies
\be
k^a\grad_a q_{ij} = 0.
\ee
This proves statement~(i).

To prove statement~(ii), first we take the directional derivative of~\eqref{eq:xigradk} along~$k^a$.  Using the Riemann tensor of pure AdS~\eqref{eq:riemann} as well as the fact that~$k^a$ is affinely-parametrized, geodesic, and orthogonal to the~$\xi_i^a$, we can show that
\be
k^a \grad_a \alpha_i = 0.
\ee
Thus the~$\alpha_i = \alpha_i(s)$ are independent of~$\lambda$.

Next, let us think of the~$\alpha_i$ as components of some one-form~$\bm{\alpha}$ on~$\Sigma$; we will show that~$\bm{\alpha}$ is closed.  Starting from~\eqref{eq:xigradk} and defining a derivative operator on~$\Sigma$ as~$D_i = \xi_i^a \grad_a$, we have
\bea
k^b D_{[i} \alpha_{j]} = k^b \xi_{[i|}^a \grad_a \alpha_{|j]} &= \xi_{[i|}^a \grad_a (\xi_{|j]}^c \grad_c k^b ) \nonumber \\
			&= \xi_{[i}^a \xi_{j]}^c \grad_a  \grad_c k^b +(\xi_{[i|}^a \grad_a \xi_{|j]}^c) \grad_c k^b \nonumber \\
			&= \frac{1}{2} \left(- \xi_i^a \xi_j^c k^d R_{acd}^{\phantom{acd}b} + \left[\xi_i,\xi_j\right]^c \grad_c k^b \right).
\eea
The~$\xi_i^a$ commute since they are coordinate basis vectors, so the second term vanishes.  The first term vanishes using~\eqref{eq:riemann} and the fact that the~$\xi_i^a$ are orthogonal to~$k^a$, and so
\be
D_{[i} \alpha_{j]} = 0.
\ee
Thus as promised,~$\bm{\alpha}$ is closed and thus exact by the Poincar\'e lemma\footnote{Here we make use of the fact that~$\Sigma$ is contractible.}; that is, there exists some scalar~$f$ on~$\Sigma$ such that
\be
\alpha_i = D_i f = \frac{\partial f}{\partial s^i}.
\ee
Then by rescaling~$\lambda \to e^{-f} \lambda$, the~$\xi_i^a$ change by
\be
\xi_i^a \to \xi_i^a + \lambda \, \alpha_i k^a,
\ee
and so~$k^b \grad_b \xi_i^a \to 0$.  This proves statement~(ii).

\subsection{Explicit Construction}
\label{subapp:explicit}

In order to provide a proof by explicit construction, we start with AdS$_{d+1}$ in Poincar\'e coordinates:
\be
ds^2 = \frac{\ell^2}{z^2}\left[-dt^2 + dz^2 + \sum_i (dx^i)^2\right],
\ee
where~$\ell$ is the AdS scale.  By exploiting the translational symmetries of these coordinates, it is easy to show that any null geodesic anchored at a fixed boundary point~$(t = \tau,x^i = 0)$ can be parametrized in terms of some affine parameter~$\lambdat$ as
\be
\label{eq:Poincaregeodesic}
\big(t(\lambdat),z(\lambdat),x^i(\lambdat)\big) = \left(-\frac{\ell^2}{\lambdat} + \tau, -\frac{\ell^2\sqrt{1-\vec{p}^2}}{\lambdat}, - \frac{\ell^2 p^i}{\lambdat}\right),
\ee
where~$p^i$ are a set of constants labeling the geodesic (which are the conserved momenta associated with the translational Killing fields~$(\partial_{x^i})^a$ normalized by the conserved energy associated with~$(\partial_t)^a$).  Taking the variables~$(\tau,\lambdat,p^i)$ as a new coordinate system, the metric takes the form
\be
ds^2 = \frac{1}{1-\vec{p}^2} \left[-(\lambdat/\ell)^2 d\tau^2 - 2 d\tau \, d\lambdat +  \ell^2\sum_{i,j} \left(\delta_{ij} + \frac{p^i p^j}{1-\vec{p}^2}\right) dp^i \, dp^j \right].
\ee
Next, we rescale~$\lambdat$ by a~$p$-dependent factor designed to ensure that the coordinate basis~$(\partial_{p^i})^a$ is parallel-transported along~$(\partial_\lambda)^a$:
\be
\lambdat = \sqrt{1-\vec{p}^2} \, \lambda.
\ee
In principle we are done, but to highlight the spatial geometry~$\Sigma$ we convert from the Cartesian~$p^i$ to spherical coordinates~$(p,\Omega)$ (with~$\Omega$ any set of coordinates on~$S^{(d-2)}$), and then take~$p = \tanh\chi$.  The metric thus becomes
\be
\label{eq:adaptedmetric}
ds^2 = -(\lambda/\ell)^2 \, d\tau^2 + 2 \left(\lambda \sinh\chi \, d\chi - \cosh\chi \, d\lambda \right)d\tau + \ell^2\left[d\chi^2 + \sinh^2\chi \, d\Omega^2_{d-2}\right].
\ee
By construction, any surface of constant~$\tau$ is a Poincar\'e horizon~$\Hcal$ of AdS, and~$\lambda$ is an affine parameter along its generators.  It is clear from the above form that the spatial geometry~$\Sigma$ of these surfaces is just hyperbolic space~$H^{(d-1)}$ and is~$\lambda$-independent, proving statement~(i).  As can be checked by explicitly computing Christoffel symbols, the coordinate basis vector field~$k^a = (\partial_\lambda)^a$ is null and geodesic, while any coordinate basis vectors on the~$H^{(d-1)}$ are parallel transported along~$k^a$.  This proves statement~(ii).

As an aside, note that this coordinate choice is highly non-unique; we have fixed the scaling of~$\lambda$ on surfaces of constant~$\tau$, but we are still free to perform the residual reparametrizations
\be
\tau \to T(\tau), \qquad \lambda \to A(\tau) \lambda + B(\tau,h^i), \qquad h^i \to H^i(\tau,h^i)
\ee
for arbitrary functions~$T$,~$A$,~$B$,~$H^i$, and where we schematically denote coordinates on~$H^{(d-1)}$ as~$h^i$.

Finally, let us note for future use that by using the standard coordinate transformations between global and Poincar\'e coordinates of AdS, we find that we can obtain the metric~\eqref{eq:adaptedmetric} from that of global AdS,
\be
\label{eq:AdSglobal}
ds^2 = -(1+r^2/\ell^2) d\ttt^2 + \frac{dr^2}{1+r^2/\ell^2} + r^2 \left(d\theta^2 + \sin^2\theta \, d\Omega_{d-2}^2\right),
\ee
via the coordinate transformations
\begin{subequations}
\label{eqs:globaltoadapted}
\bea
&\tan(\ttt/\ell) = \frac{\ell \cosh\chi - \lambda \tau/\ell}{(\lambda/2)(\tau^2/\ell^2 - 1) - \tau\cosh\chi}, \\
&\tan\theta = \frac{\ell\sinh\chi}{(\lambda/2)(\tau^2/\ell^2 + 1) - \tau\cosh\chi}, \\
&r^2 = \ell^2 \sinh^2\chi + \left(\frac{\lambda}{2}\left(\frac{\tau^2}{\ell^2} + 1\right) - \tau\cosh\chi\right)^2.
\eea
\end{subequations}

\section{Linearized Averaged Null Curvature}
\label{app:ANCC}

In this Appendix, we derive expression~\eqref{eq:linANCC} for the linearized averaged null curvature.  First, note that along any null geodesic~$\gamma$ of the background AdS we have
\be
\label{eq:tempANCC}
\delta \left(\intg R_{ab} k^a k^b \, d\lambda\right) = \intg \left(\delta R_{ab} k^a k^b + 2R_{ab} k^a \delta k^b \right) d\lambda.
\ee
(The linearization commutes with the integration if we take~$\lambda$ to be an affine parameter along both the unperturbed and perturbed geodesic; this is allowed by the residual gauge freedom to shift and rescale~$\lambda$.)  Now, the second term can be evaluated in a straightforward manner by recalling that~$\delta k^a = k^b \grad_b \eta^a$, and using~\eqref{eq:constraint} and the Ricci tensor of AdS~$R_{ab} = -(d/\ell^2) g_{ab}$:
\be
\label{eq:ANCCterm2}
\intg 2R_{ab} k^a \delta k^b \, d\lambda = \frac{d}{\ell^2}\intg \delta g_{ab} k^a k^b \, d\lambda.
\ee

The first term of~\eqref{eq:tempANCC} requires a bit more work.  First, recall that under an arbitrary metric perturbation the Ricci tensor changes as
\be
\delta R_{ab} = \frac{1}{2}\big(\Delta_L \delta g_{ab} + \grad_a \grad^c \delta g_{cb} + \grad_b \grad^c \delta g_{ca} - \grad_a \grad_b \delta g \big),
\ee
where~$\delta g \equiv g^{ab} \delta g_{ab}$ and the Lichnerowicz operator acting on symmetric, rank-two tensors is
\be
\Delta_L \delta g_{ab} \equiv -\grad^c \grad_c \delta g_{ab} - 2R_{acbd} \delta g^{cd} + R^c_{\phantom{c}a} \delta g_{cb} + R^c_{\phantom{c}a} \delta g_{cb}.
\ee
Using these expressions and combining with~\eqref{eq:ANCCterm2}, we find that~\eqref{eq:tempANCC} becomes
\be
\delta \left( \intg R_{ab} k^a k^b \, d\lambda \right) = \intg \left(-\frac{1}{2} \, \grad^c \grad_c \delta g_{ab} + \frac{1}{\ell^2} \delta g_{ab}\right) k^a k^b \, d\lambda,
\ee
where we have liberally integrated by parts (using the falloffs~\eqref{eq:falloffs}).

Next, we again consider the Poincar\'e horizon~$\Hcal$ to which~$\gamma$ belongs and we decompose the Laplacian in terms of the basis~$(k^a, l^a, \xi_i^a)$:
\be
\grad^c \grad_c = g^{cd} \grad_c \grad_d = \left(2k^{(c} l^{d)} + \sum_{i,j} q^{ij} \xi_i^c \xi_j^d \right) \grad_c \grad_d.
\ee
Then after some more manipulations and integrations by parts, we find
\be
\label{eq:intermediate}
\delta \left( \intg R_{ab} k^a k^b \, d\lambda \right) = -\frac{1}{2} \sum_{i,j} q^{ij} \left[\frac{\partial}{\partial s^i} D_j I - \intg (\xi_i^c \grad_c \xi_j^d)\grad_d (\delta g_{ab} k^a k^b) d\lambda \right],
\ee
where~$I$ is as defined in~\eqref{eq:averagedclosing}.  Now, the basis vectors~$\xi_i^a$ live in~$\Hcal$, and therefore so do the vectors~$\xi_i^a \grad_a \xi_j^b$.  The latter must therefore admit an expansion of the form
\be
\label{eq:xigradxi}
\xi_i^a \grad_a \xi_j^b = \sum_{k} a^k_{ij} \xi_k^b + b_{ij} k^b
\ee
for some connection coefficients~$a^k_{ij}$,~$b_{ij}$.  Contracting~\eqref{eq:xigradxi} with~$\sum_n q^{mn} \xi_n^a$, we find that the~$a^k_{ij}$ are just the Christoffel symbols of~$\Sigma$:
\be
a^k_{ij} = \sum_m q^{km} (\xi_m)_b \xi_i^a \grad_a \xi_j^b = \Gamma^k_{ij}.
\ee
These are computed from~$q_{ij}$, and are thus independent of~$\lambda$.  To find the~$\lambda$-dependence of the~$b_{ij}$, we take the covariant derivative along~$k^a$ of~\eqref{eq:xigradxi}; then once again using the Riemann tensor~\eqref{eq:riemann} as well as the facts that~$\Gamma^k_{ij}$,~$k^a$, and~$\xi_i^a$ are all covariantly constant along~$k^a$, we obtain
\be
\left(k^a \grad_a b_{ij}\right) k^b = -\frac{1}{\ell^2} \, q_{ij} \, k^b.
\ee
Thus
\be
b_{ij} = -\frac{1}{\ell^2} \, \lambda \, q_{ij} + \beta_{ij}
\ee
for some functions~$\beta_{ij}$ which are independent of~$\lambda$ (in fact, they are related to the residual freedom to shift~$\lambda$, so they must be proportional to~$q_{ij}$, and so if desired can be made to vanish by an appropriate shift in~$\lambda$).  Then using this expression and that for the~$a^k_{ij}$, we plug~\eqref{eq:xigradxi} into~\eqref{eq:intermediate} and as promised finally obtain (after one more integration by parts)~\eqref{eq:linANCC}.


\bibliographystyle{jhep}
\bibliography{all}

\end{document}